\documentclass[aps,pre,floatfix,twocolumn,showpacs,amsmath,amssymb,superscriptaddress,titlepage,natbib]{revtex4-1}
\usepackage{graphicx,times}
\usepackage{subfigure}
\usepackage{amssymb,amsmath}
\usepackage{multirow}
\usepackage{color}
\usepackage{epstopdf}
\usepackage{bm}% bold math
\usepackage{longtable}
\usepackage{natbib}
\usepackage{lipsum}
\usepackage[colorlinks=true, urlcolor=blue, linkcolor=red, citecolor=blue, pdftex]{hyperref}
\hypersetup{pdfstartview={FitV}, pdfpagemode = UseNone, pdfpagelayout=SinglePage, bookmarksnumbered={true}}

\begin{document}

\title{Contrariety and Inhibition enhance synchronization in a  small-world network of phase oscillators}

\author{Tayebe Nikfard}
\affiliation{ Department of Physics, Isfahan University of Technology, Isfahan 84156-83111, Iran}
%nikfardtayebeh@yahoo.com

\author{Yahya Hematyar Tabatabaei}
\affiliation{ Department of Physics, Isfahan University of Technology, Isfahan 84156-83111, Iran}
%y.hematyar@ph.iut.ac.ir

\author{ Farhad Shahbazi} \email{shahbazi@iut.ac.ir }
\affiliation{ Department of Physics, Isfahan University of Technology, Isfahan 84156-83111, Iran}

%\author{Reihaneh Kouhi Esfahani}
%\affiliation{ Department of Physics, Isfahan University of Technology, Isfahan 84156-83111, Iran}

\begin{abstract}
We numerically study the Kuramoto model's synchronization consisting of the two groups of conformist-contrarian and excitatory-inhibitory phase oscillators with equal intrinsic frequency. We consider random and small-world (SW) topologies for the connectivity network of the oscillators. In random networks, regardless of the contrarian to conformist connection strength ratio, we found a crossover from the $\pi$-state to the blurred $\pi$-state and then a continuous transition to the incoherent state by increasing the fraction of contrarians. However, for the excitatory-inhibitory model in a random network, we found that for all the values of the fraction of inhibitors, the two groups remain in-phase, and the transition point of fully synchronized to incoherent state reduced by strengthening the ratio of inhibitory to excitatory links. In the SW networks, we found that the order parameters for both models have not monotonic behavior in terms of the fraction of contrarians and inhibitors. Up to the optimal fraction of contrarians inhibitors, the synchronization rises by introducing the contrarians and inhibitors and then falls. We discuss that this non-monotonic behavior in synchronization is due to the weakening of the defects formed already in conformists and excitatory agents in SW networks. 
\end{abstract}


\maketitle

\section{Introduction}

The Kuramoto model originally introduced by Y. Kuramoto~\cite{kuramoto1975self,kuramoto2012chemical} is a simple and 
mathematically tractable model, showing the synchronization transition in an ensemble of mutually interacting phase-oscillators (rotors).  
Synchronization is ubiquitous such as flashing of fireflies, flocking of birds and fishes, the simultaneous firing of brain neurons, and heart cells~\cite{winfree2001geometry,pikovsky2003synchronization, manrubia2004emergence, balanov2008synchronization,strogatz2012sync}. 

While Kuramoto's motivation for introducing his model was to find an exactly solvable model for the transition from desynchronized to synchronized state, however, this model found many applications in physical, chemical, and biological systems \cite{acebron2005kuramoto}.
The Kuramoto model and its variants have been used to model the opinion formation
 dynamics in a society~\cite{pluchino2005changing,pluchino2006compromise,pluchino2006opinion,hong2011conformists,hong2011kuramoto}. In this context, the phase of each individual corresponds to its belief  and synchronization is equivalent to the formation of consensus in society.

In the original Kuramoto model, the rotors globally coupled to each other sinusoidally  with uniform coupling constant. The exact solution of this model indicates a continuous synchronization at a critical coupling~\cite{acebron2005kuramoto}.
Recent developments in complex network science attracted the researchers to the study of synchronization in general ~\cite{arenas2008synchronization}, and the Kuramoto model with local interactions in complex networks~\cite{rodrigues2016kuramoto}.
For instance, the Kuramoto model has been studied on the regular~\cite{wiley2006size,delabays2017size} and WS networks~\cite{gade2000synchronous,watts2001small,hong2002synchronization, barahona2002synchronization,  wang2002synchronization,esfahani2012noise}. 
In this regard, some impressive results are forming patterns of synchrony in regular and Watts-Strogatz (WS) networks for a group of identical phase oscillators. In a regular ring shape network, the patterns are helical, topologically distinct, and characterized by integer winding numbers~\cite{wiley2006size,delabays2017size}. Rewiring the ring's links with a small probability between $0.005$ and $0.05$ converts the regular lattice to the WS network, which has the small-world properties of small mean shortest path and high clustering~\cite{watts1998collective, watts2001small}. This process turns the helical patterns into several isolated defects or deviated helical patterns for small and large winding numbers, respectively~\cite{esfahani2012noise}. In isolated defects, the phase difference between their center and surroundings varies continuously fro $0$ to $\pi$ by going away from the center. 

Kouhi et. al found that adding a spatially uncorrelated white noise to the Kuramoto model could destroy theses defects and make a more homogenous phase texture with higher phase-synchrony, that is, the {\em stochastic synchronization}~\cite{esfahani2012noise}. 
Moreover, they showed that the significant rewiring probabilities larger than  $0.15$, destroys the defects and so set the system in a full synchronization state~\cite{esfahani2012noise}.    

To make the Kuramoto model closer to reality, the generalization of the model with random pairwise coupling with both signs was investigated~\cite{daido1992quasientrainment,daido2000algebraic,stiller1998dynamics}. While the positive coupling encourages that the phases of interacting oscillators to converge an in-phase state, the negative coupling forces them to align in anti-phase ($\pi$-difference) configuration. Initially, a glassy behavior was
 claimed for this model~\cite{daido1992quasientrainment}, however, later investigations shed doubt on the existence and properties of such a state~\cite{daido2000algebraic,stiller1998dynamics}.  
 
Hong and Strogatz introduced a simpler version of the model with mixed-coupling in which the oscillators are divided into two groups of conformists and contrarians~\cite{hong2011conformists,hong2011kuramoto}. In this model, the conformists have positive coupling with the rest of oscillators, so tend to be in line with the dominant rhythm of the population; however, the contrarians have negative coupling with the rest and are prone to move in the opposite direction of population. Hong and Strogatz investigated the model with both with~\cite{hong2011conformists} and without~\cite{hong2011kuramoto} intrinsic angular-velocity distribution and found a rich phase diagram for this model in terms of the fraction of conformists. The phase diagram includes the desynchronized, anti-phase locked state between the conformists and contrarians, and the traveling waves in which the average angular velocity is different from the mean intrinsic angular-velocity distribution. Surprisingly, they found that identical phase-oscillators' model has a richer phase diagram.

Motivated by Hong and Strogatz, in this work, we study the Kuramoto model of two groups of conformist and contrarian phase-oscillators, which are identical in terms of their intrinsic angular-velocities, in a WS network. Moreover, we also investigate a Kuramoto model of excitatory and inhibitor phase-oscillators in the WS network, in which a given rotor is  coupled positively to its excitatory and negatively to its inhibitor neighbors.

The paper is organized as the following. In section~\ref{model}, we define the model and the numerical methods of quantifying the synchronization. Section~\ref{results} represents the results and  discussion and section~\ref{conclusion} is devoted to the concluding remarks.

\section{model and method}
\label{model}

In this work, we study two Kuramoto models in random and Watts-Strogatz small-world networks. The two models are i) The conformist-contrarian model and ii) the excitatory-inhibitory model.  
Each model consists of $N$ phase oscillators (rotors), divided into two groups (conformist-contrarian and excitatory-inhibitory). The rotors occupy the vertices of a network, and each interacts with its neighbors through a sinusoidal coupling whose argument is their phase difference.  
However, the coupling is not symmetric. 

\subsection{Conformist-Contrarian model}
The conformist-contrarian model is given by a set of coupled first-order differential equations, as:
% mutual coupling that is the Kuramoto coupling given by the sine of their phase difference. Then the model is given by the following set of coupled differential equations~\cite{kuramoto1975self}:
%-------------------------------------------------------------------------------------------------------------------------------------
\begin{equation}
\label{cc1}
\frac{{d\theta}_{i}^{s}}{dt}=\omega_0+\frac{\lambda_i^s}{k_i}\sum_{j=1}^{N}{a_{ij}\sin(\theta_j-\theta_i^s)},\hspace{1cm}    i=1,...,N; 
\end{equation}
%---------------------------------------------------------------------------------------------------------------------------------------
where $\theta_i$ denotes the phase of the oscillator sitting at node $i$, $ \omega_0$ is the intrinsic frequency of the oscillators and considered equal for all of them. $a_{ij}$ denotes the elements of the adjacency matrix (i.e. $a_{ij}=1$ if $i$ and $j$ are connected and $a_{ij}=0$ otherwise) and $k_i$ is the degree of node $i$. $\lambda^{s}$, where $s={\rm conformist, contrarian}$), is the coupling constant, which is positive for the conformist and negative for the contrarians. Indeed, a contrarian with positive coupling inclines to align its phase with its neighbors' phases, while a contrarian tries to direct its phase as far as possible to its neighbors. 

The intrinsic frequency, $\omega_0$), can always be set to zero by moving to a rotating frame with the angular velocity $\omega_0$, i.e. $\theta'(t)=\theta(t)-\omega_0 t$. Moreover, by defining the dimensionless time $\tau=\lambda^{\rm conf} t$, Eq.\eqref{cc1} converts to

\begin{eqnarray}
\label{cc2}
&&\frac{{d\theta'}_{i}^{\rm conf}}{d\tau}=\frac{1}{k_i}\sum_{j=1}^{N}{a_{ij}\sin(\theta_j^{'}-\theta_i^{'\rm conf})},\nonumber\\  
&&\frac{{d\theta'}_{i}^{\rm cont}}{d\tau}=-\frac{Q}{k_i}\sum_{j=1}^{N}{a_{ij}\sin(\theta_j^{'}-\theta_i^{'\rm  cont})},  
\end{eqnarray}
where we assumed $\lambda^{\rm cont}=-Q\lambda^{\rm conf}$, with $Q>0$ and  $\lambda^{\rm conf}>0$ ({\rm cont} and {\rm conf} stand for contrarian and conformist, respectively))

\subsection{Excitatory-Inhibitory model}
Now consider a system consisting of two groups of excitatory and inhibitory phase oscillators. Each oscillator receives a positive input from its excitatory and negative input from its inhibitory neighbors in this system. Therefore the coupling $\lambda_j^s$ has to inserted inside the sum, then we find 

\begin{equation}
\label{ei1}
\frac{{d\theta}_{i}}{dt}=\omega_0+\frac{1}{k_i}\sum_{j=1}^{N}{a_{ij}\lambda_j^s\sin(\theta_j^{s}-\theta_i)},\hspace{1cm}    i=1,...,N; 
\end{equation}
in which we assume $\lambda_i^{\rm inhib}=-Q\lambda_i^{\rm excit}$ with  $Q>0$ and $\lambda^{\rm excit}>0$. ({\rm inhib} and {\rm excit} stand for inhibitory and excitatory, respectively). Similarly, after moving to the rotating frame with angular velocity $\omega_0$ and rescaling the time variable as 
$\tau=\lambda^{\rm excit} t$, we get

\begin{equation}
\label{ei2}
\frac{{d\theta'}_{i}}{d\tau}=\frac{1}{k_i}\sum_{j\in \rm{excit}}{a_{ij}\sin(\theta_j^{'}-\theta_i^{'})}-\frac{Q}{k_i}
\sum_{j\in \rm {inhib}}{a_{ij}\sin(\theta_j^{'}-\theta_i^{'})}.
\end{equation}
 
\subsection{method}

To obtain the time evolution of the phase of the oscillators, we use the fourth-order Runge-Kutta method for integrating the sets of equations~\eqref{cc2} and \eqref{ei2}. The integration time step is set to $d\tau=0.1$ and the initial phase distribution is taken from a box distribution in the interval $[-\pi,\pi]$.
The global synchrony among the oscillators at any time can be measured by the Kuramoto order parameter defined as
%-----------------------------------------------------------------------------------------------------------------------------
\begin{equation}
\label{r}
{\bf r}(\tau)=\frac{1}{N}\sum_{j=1}^{N}\exp{(i\theta_j(\tau))}.
\end{equation}
%-----------------------------------------------------------------------------------------------------------------------
In the stationary state we define the long time averaged order parameter as $r_{\infty}$:
%------------------------------------------------------------------------------------------------------------------------
\begin{equation}
\label{rstat}
    {\bf r}_{\infty}= \lim_{\Delta \tau \to \infty}\frac{1}{\Delta \tau}\int^{\tau_s +\Delta\tau}_{\tau_s}r(\tau) d\tau,
\end{equation}
%------------------------------------------------------------------------------------------------------------------------
in which $\tau_s$ is the time of reaching a stationary state.
The magnitude of ${\bf r}_{\infty}$ shown by ${r}_{\infty}$ is a real number between $0$ and $1$. 
$r_{\infty}=0$ indicates a disordered or a phase-locked state with regular phase lag, $0<r_{\infty}<1$ indicates a partially synchronized state, and $r_{\infty}=1$ shows the full synchrony in the system. 

We also define a partial order parameter for the two groups of oscillators as:

%-------------------------------------------------------------------------------------------------------------------------
\begin{equation}
\label{r_partial}
{\bf r}_a(\tau)=r_a(\tau)\exp(i\Theta_a(\tau))=\frac{1}{N_a}\sum_{j\in a}\exp{(i\theta^{a}_j(\tau))}.~a=1,2
\end{equation}
%-------------------------------------------------------------------------------------------------------------------------
where $a=1$ refers to conformist and excitatory and $a=2$ to contrarian and inhibitory oscillators. $r_a$ and $\Theta_a$ are the magnitude and phase of the partial order parameter. 
For each model the fraction of contrarian ($\frac{N_{\bf ct}}{N}$) and inhibitory ($\frac{N_{\bf in}}{N}$) oscillators is denoted by $p$.

It can be easily seen that equations \eqref{r} and \eqref{r_partial} give rise to:

%-----------------------------------------------------------------------------------------------------------------
\begin{equation}
\label{r_partial2}
{\bf r}(\tau)=(1-p) {\bf r}_{1}(\tau)+p {\bf r}_{2}(\tau),
\end{equation}
%-----------------------------------------------------------------------------------------------------------------
from which one fined the following expression for the magnitude of the total order parameter in terms of the magnitude of partial order parameters and their phase difference
%-----------------------------------------------------------------------------------------------------------------
\begin{equation}
\label{r_partial3}
{ r}=\sqrt{(1-p)^{2} { r}^{2}_{1}+p^{2} {r}^{2}_{2}+2p(1-p)r_{1}r_{2}\cos(\Theta_{1}-\Theta_{2})}.
\end{equation}
%-----------------------------------------------------------------------------------------------------------------

The Kuramoto  order parameter is a measure of global synchronization in the system. To gain insight into the local coherence  in the stationary state, we calculate the pairwise correlation matrix $D$ \cite{gomez2007paths}:
%-------------------------------------------------------------------------------------------------------------------------------
\begin{equation}
\label{D-matrix}
    D_{ij}= \lim_{\Delta \tau \to \infty}\frac{1}{\Delta \tau}\int^{\tau_s+\Delta\tau}_{\tau_s}\cos(\theta_i(\tau)-\theta_j(\tau))dt.
\end{equation}
%--------------------------------------------------------------------------------------------------------------------------------
%where the correlation matrix element $D_{ij}$ is a coherency between the pair of oscillators $i$ and $j$. 
The matrix elements $D_{ij}$ take a value in the interval $[-1,1]$. $D_{ij}=1$ denotes full synchrony between oscillators $i$ and $j$, while $D_{ij}=-1$, represents an anti-phase state (i.e.$|\theta_i-\theta_j| = \pi$).

The phase diagram of the conformist-contrarian model, with identical intrinsic frequency, has been obtained by Hong and Strogatz~\cite{hong2011kuramoto}. This phase diagram includes four phases \\
a) {\em incoherent} phase where both partial order parameters are zero ($r_{1}=r_{2}=0$),\\
b) {\em blurred} $\pi$-state, where the conformists and contrarians are partially ordered ($r_{1}, r_{2} < 1$), but there is $\pi$ difference between the phase of their order parameters 
$|\Theta_{1}-\Theta_{2}|=\pi$. The incoherent and blurred $\pi$-states coexists together for some range of fraction of contrarians. \\
c) {\em traveling wave} state, where the conformist are completely ordered $r_{1}=1$, but the conformist are partially ordered $r_{2}<1$, in this case, the absolute phase difference of these order parameters could be less than $\pi$, and \\
d) the $\pi$-state, where both conformists and contrarians are fully synchronized $r_{1}=r_{2}=1$ and their phase difference is equal to $\pi$ ($|\Theta_{1}-\Theta_{2}|=\pi$).

In next section we proceed to investigate the phase diagram of both conformist-contrarian and excitatory-inhibitory models in random and small-world networks. The networks used in this study are created by Watts-Strogatz (WS) algorithm~\cite{watts1998collective,watts2001small}. To build an SW network, starting from a ring with a given degree of $k$ for each node, we rewire the links with the probability $0.03$, and for the random network, we rewire them with probability $1$.

%============================
\section{results and discussion}
\label{results}

%%%%%%%%%%%%%%%%%%%%%%%%%%%%%%%%%%%%%%%%%%%%%%%%%%%%%%%%%%fig.1
\begin{figure}[t]
    \centering
   \subfigure[\label{figure label}] {\includegraphics[width=0.8\columnwidth]{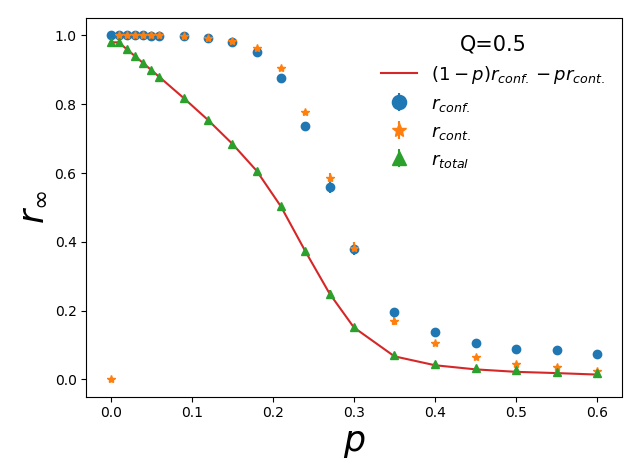}}
    \subfigure[\label{figure label}] {\includegraphics[width=0.8\columnwidth]{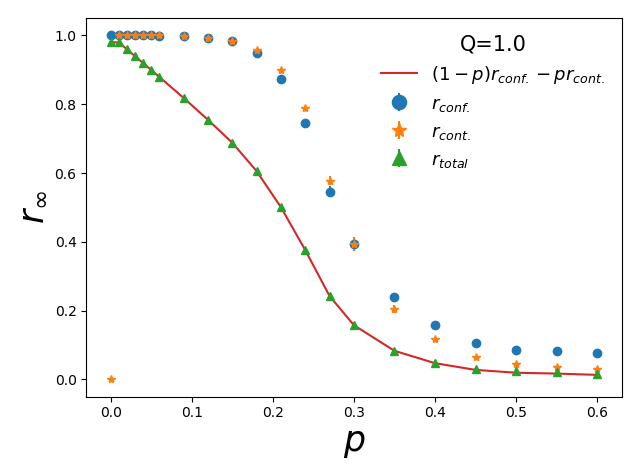}}
    \subfigure[\label{figure label}] {\includegraphics[width=0.8\columnwidth]{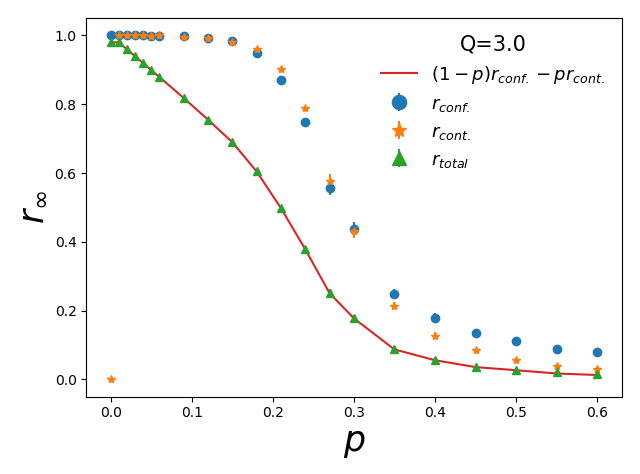}}
    \caption{(Color online) Stationary total and partial order parameters verses  fraction of contrarians for the conformist-contrarian model (Eq.\eqref{cc2}) 
    for (a) $Q=0.5$, (b) $Q=1.0$ and (c) $Q=3.0$ for a random network of $N = 1000$ oscillators and mean degree $<k>=10$. The error bars indicate the standard error of mean (SEM)}
    \label{fig1:r-cc}
\end{figure}
%%%%%%%%%%%%%%%%%%%%%%%%%%%%%%%%%%%%%%%%%%%%%%%%%%%%%%%%%%%%%

%%%%%%%%%%%%%%%%%%%%%%%%%%%%%%%%%%%%%%%%%%%%%%%%%%%%%%%%%fig.2
\begin{figure}[t]
    \centering
   \subfigure[\label{figure label}] {\includegraphics[width=0.8\columnwidth]{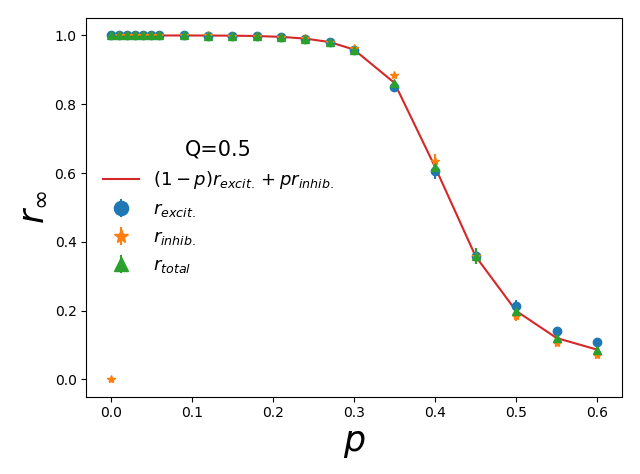}}
    \subfigure[\label{figure label}] {\includegraphics[width=0.8\columnwidth]{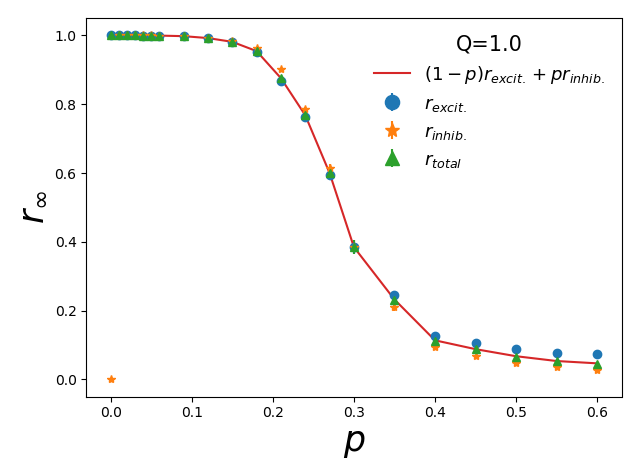}}
    \subfigure[\label{figure label}] {\includegraphics[width=0.8\columnwidth]{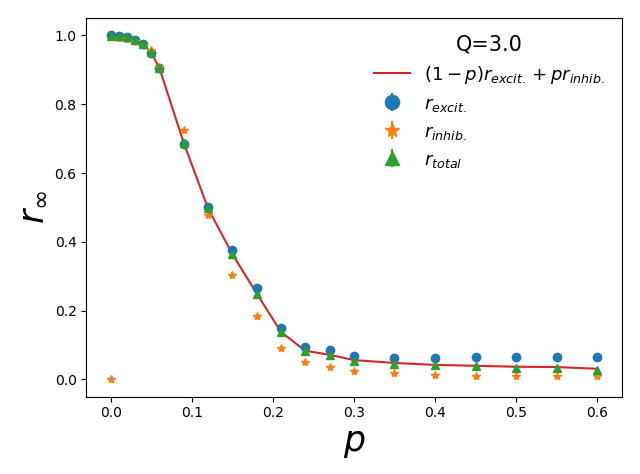}}
     \caption{(Color online) Stationary total and partial order parameters verses  fraction of inhibitory oscillators for the excitatory-inhibitory  
                   (Eq.\eqref{ei2}) model  for (a) $Q=0.5$, (b) $Q=1.0$ and (c) $Q=3.0$ for a random network of $N = 1000$ oscillators and mean degree  $<k>=10$. The error bars indicate the standard error of mean (SEM)}
    \label{fig2:r-ei}
\end{figure}

Since we did not find remarkable size dependence for the networks larger than $N=200$, we performed all the simulations on the size $N=1000$. Moreover, since we are interested in applying the models in real networks that are mostly sparse,
we consider the mean degree of $10$. We check that, as far as the network is sparse, the results remain unchanged. In the following, we report the results of both models in random and then in small-world networks. 

\subsection{Random network}
 
% \subsubsection{conformist-contrarian model}     
In random networks for the conformist-contrarian model, the simulations typically reach the stationary state up to $\sim 10000$ time steps.  
Figure \ref{fig1:r-cc} represents the variation of the magnitude of long-time averaged total ($r_{\infty}$) and partial order parameters ($r_{\rm conf}$ and $r_{\rm cont}$), versus the fraction of contrarians ($p$) for the conformist-contrarian model with $Q=0.5, 1.0, 3.0$. Each data is obtained by averaging over $100$ independent distribution of contrarians. 

Our calculations indicate that for all ranges of $p$, the phase difference between two partial order parameters is equal to $\pi$, i.e., the system is in $\pi$ or blurred $\pi$-state for all values of $p$. 
 In this case, Eq.\eqref{r_partial3} results in the following relation for the stationary total order parameter in terms of the partial order parameters

%-----------------------------------------------------------------------------------------------------------------
\begin{equation}
\label{r-cc-rand}
r_{\infty}=(1-p) r_{\rm conf}-p r_{\rm cont}.
\end{equation}
%-----------------------------------------------------------------------------------------------------------------
The solid line in figure~\ref{fig1:r-cc} is the plot of Eq.~\ref{r-cc-rand} which completely lies
on the total order parameter data.  
This figure shows, regardless of the value of $Q$, the magnitude of partial order parameters is near unity for $p\lesssim 0.12$, giving rise to a linear dependence of $r_{\infty}$ versus $p$ with the slope $-2$ in this range of $p$. It means that the stable state of the system in this region is $\pi$-state. However, for $p > 0.12$, conformists and contrarians' partial order parameters are less than unity; indicating that they have more freedom to swing their phase,  then the system is the blurred $\pi$-state. Our results denote that the traveling wave states are not present in a random network with sparse connectivity, unlike the all-to-all network. 
  
Moreover, the variation of the order parameters in terms of the fraction of contrarians does not depend on $Q$. Such independence if the stationary results to the parameter $Q$, can be easily justified by absorbing $Q$ into the dimensionless time parameter $\tau$ in Eq.~\eqref{cc2}, which drops $Q$ from the right-hand side of this equation. It means that $ Q $ only affects the contrarians' stationary time scale; hence, the results in the stationary state do not depend on $ Q $.  

Indeed, we find a crossover from the $\pi$-state to the blurred 
$\pi$-state at the same $p$, for all values of $Q$. This result is in sharp contrast with the complete network;
A traveling wave (TW) state mediates the $\pi$-state and blurred $\pi$-state. The transition between the $\pi$-state and TW is discontinuous, while TW continuously connects to the blurred $\pi$-state~\cite{hong2011kuramoto}.

%\subsubsection{excitatory-inhibitory model}

The stationary time for the excitatory-inhibitory model is more considerable and could be up to 
$\sim35000$ time steps. 
Figure~\ref{fig2:r-ei} displays the dependence of the magnitude of stationary total order parameter ($r_{\infty}$) and partial order parameters ($r_{\rm excit}$ and $r_{\rm inhib}$), to the fraction of inhibitory oscillators ($p$) for the this model with $Q=0.5, 1.0, 3.0$. Each data is obtained by averaging over $100$ independent distribution of inhibitors. 

We found that the excitatory and inhibitory oscillators have the same phase for all $p$ and $ Q $ for this model; which can be verified by the exact overlap of the equation

%-----------------------------------------------------------------------------------------------------------------
\begin{equation}
\label{r-ei-rand}
r_{\infty}=(1-p) r_{\rm excit}+p r_{\rm inhib},
\end{equation}
%-----------------------------------------------------------------------------------------------------------------
with the total order parameter, illustrated  in figure~\ref{fig2:r-ei}.  

Interestingly, the system remains in a fully synchronized state (up to $p\sim 0.3$) for $Q=0.5$, and the range of full synchrony decreases by increasing $Q$.  Indeed, at small enough $p$, the minority inhibitors synchronize with each other through their neighboring majority excitatory oscillators. 
The larger $Q$  also reduces the transition point to the incoherent state. This result is not surprising; as expected, the strengthening of the inhibitory links leads to the faster vanishing of the synchronized state.  

%%%%%%%%%%%%%%%%%%%%%%%%%%%%%%%%%%%%%%%%%%%%%%%%%%%%%%%%%%fig.3
\begin{figure}[t]
    \centering
   \subfigure[\label{figure label}] {\includegraphics[width=0.8\columnwidth]{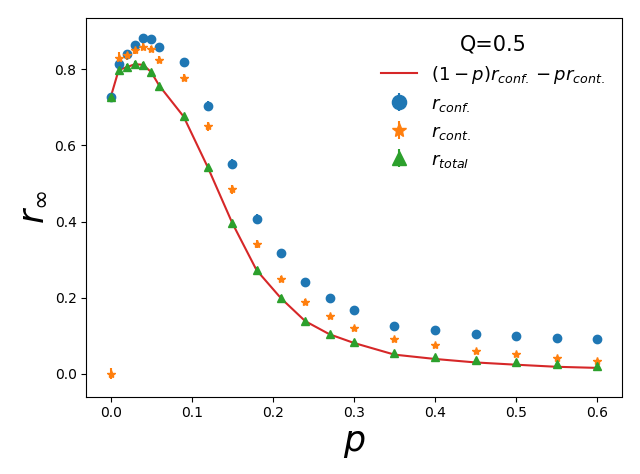}}
    \subfigure[\label{figure label}] {\includegraphics[width=0.8\columnwidth]{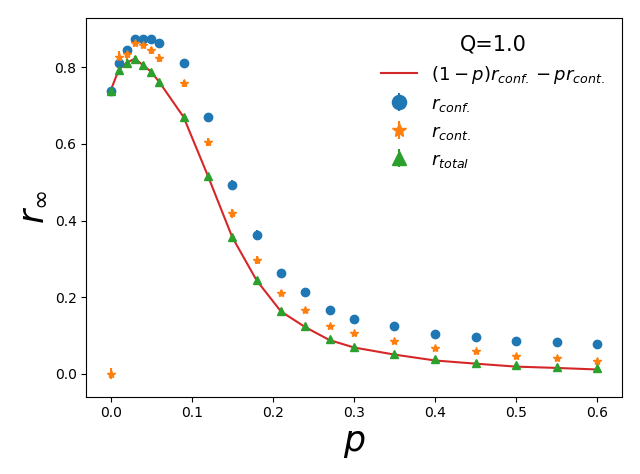}}
    \subfigure[\label{figure label}] {\includegraphics[width=0.8\columnwidth]{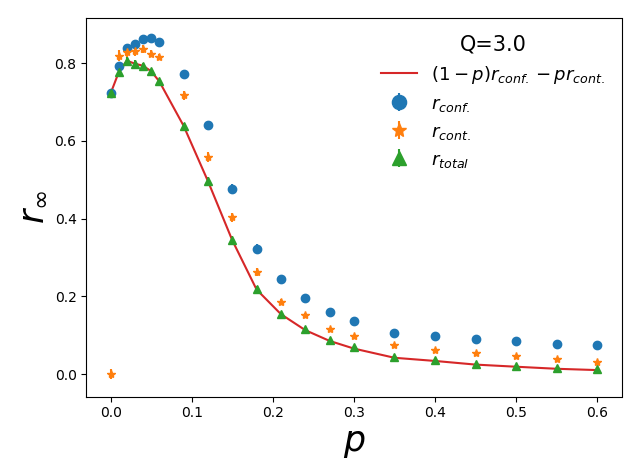}}
    \caption{(Color online) Stationary total and partial order parameters verses  fraction of contrarians for the conformist-contrarian model (Eq.\eqref{cc2})  for (a) $Q=0.5$, (b) $Q=1.0$ and (c) $Q=3.0$ for a small-world of $N = 1000$ oscillators and mean degree $<k>=10$.The error bars indicate the standard error of mean (SEM)}
    \label{fig3:r-cc}
\end{figure}
%%%%%%%%%%%%%%%%%%%%%%%%%%%%%%%%%%%%%%%%%%%%%%%%%%%%%%%%%%%%%

%%%%%%%%%%%%%%%%%%%%%%%%%%%%%%%%%%%%%%%%%%%%%%%%%%%%%%%%%fig.4
\begin{figure}[t]
    \centering
   \subfigure[\label{figure label}] {\includegraphics[width=0.8\columnwidth]{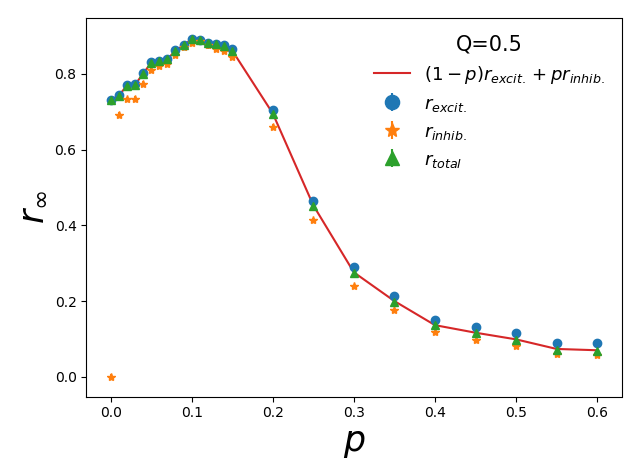}}
    \subfigure[\label{figure label}] {\includegraphics[width=0.8\columnwidth]{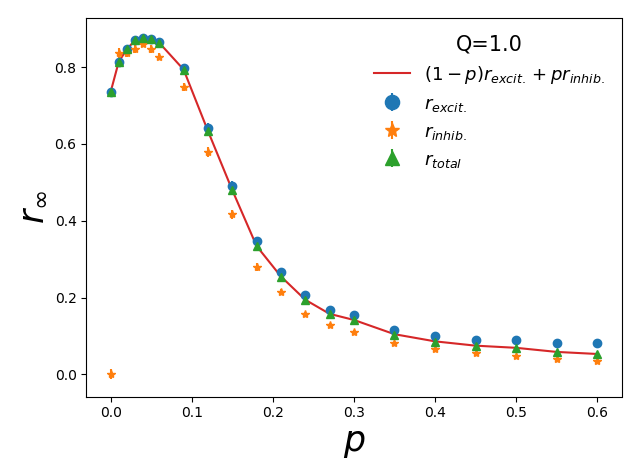}}
    \subfigure[\label{figure label}] {\includegraphics[width=0.8\columnwidth]{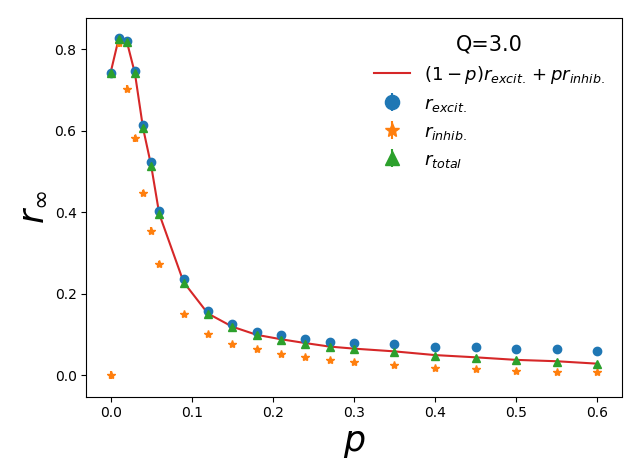}}
     \caption{(Color online) Stationary total and partial order parameters verses  fraction of inhibitory oscillators for the excitatory-inhibitory  
                   (Eq.\eqref{ei2}) model  for (a) $Q=0.5$, (b) $Q=1.0$ and (c) $Q=3.0$ for a small-world of $N = 1000$ oscillators and mean degree   $<k>=10$. The error bars indicate the standard error of mean (SEM)}
    \label{fig4:r-ei}
\end{figure}
%%%%%%%%%%%%%%%%%%%%%%%%%%%%%%%%%%%%%%%%%%%%%%%%%%%%%%%%%%%%%

%%%%%%%%%%%%%%%%%%%%%%%%%%%%%%%%%%%%%%%%%%%%%%%%%%%%%%%%%%fig.5
\begin{figure*}[]
\centering
\subfigure[\label{figure label}]{\includegraphics[width=0.7\columnwidth]{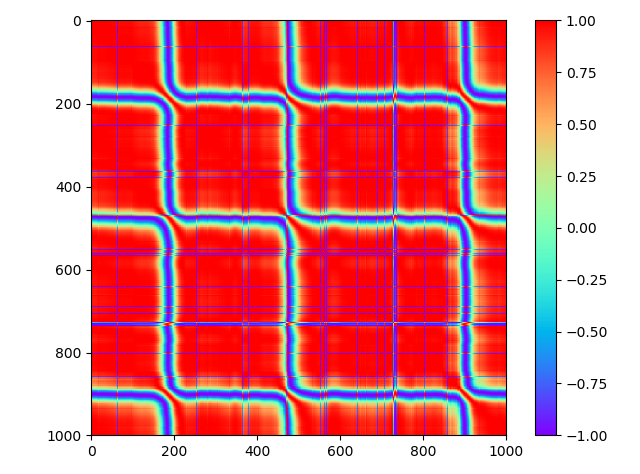}}
\subfigure[\label{figure label}]{\includegraphics[width=0.7\columnwidth]{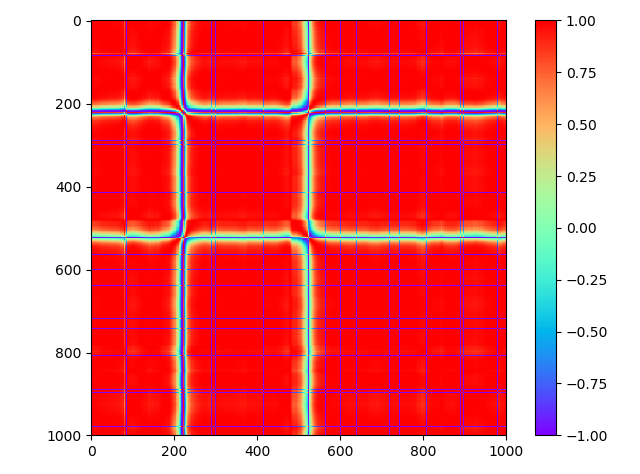}}
\subfigure[\label{figure label}]{\includegraphics[width=0.7\columnwidth]{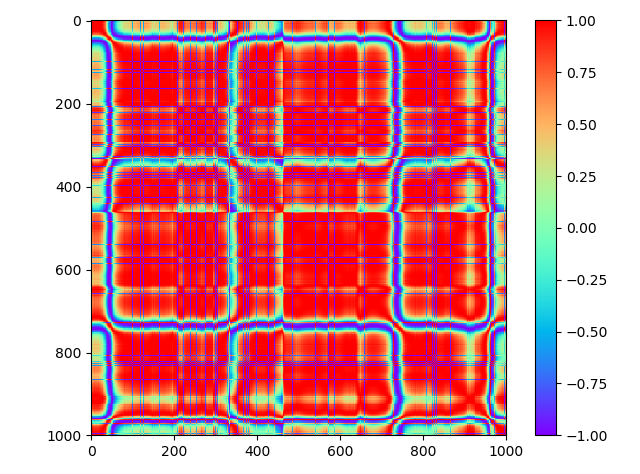}}
\subfigure[\label{figure label}]{\includegraphics[width=0.7\columnwidth]{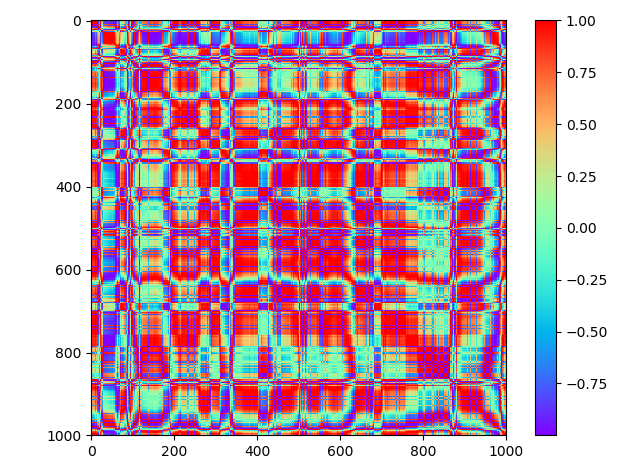}}
\caption{(Color online) Stationary density plot of the correlation matrix elements corresponding to a single realization of the conformist-contrarian model (Eq.\eqref{cc2}), with $Q=1.0$, in an SW network, for (a) $p=0.03$, (b) $p=0.05$, (c) $p=0.09$ and (d) $p=0.18$.}
\label{fig5:D-cc}
\end{figure*}
%%%%%%%%%%%%%%%%%%%%%%%%%%%%%%%%%%%%%%%%%%%%%%%%%%%%%%%%%%

%%%%%%%%%%%%%%%%%%%%%%%%%%%%%%%%%%%%%%%%%%%%%%%%%%%%%%%%%%fig.6
\begin{figure*}[]
\centering
\subfigure[\label{figure label}]{\includegraphics[width=0.7\columnwidth]{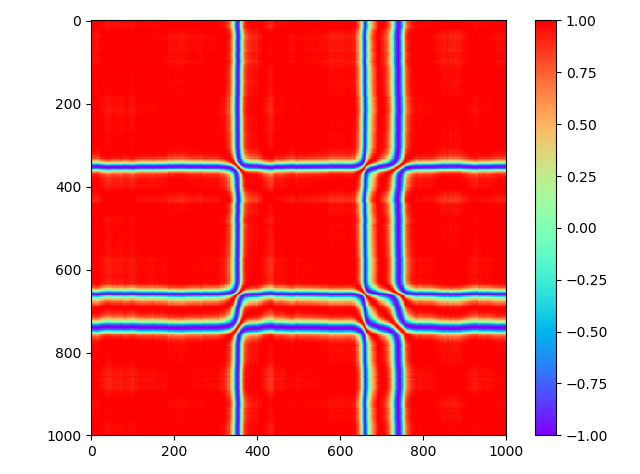}}
\subfigure[\label{figure label}]{\includegraphics[width=0.7\columnwidth]{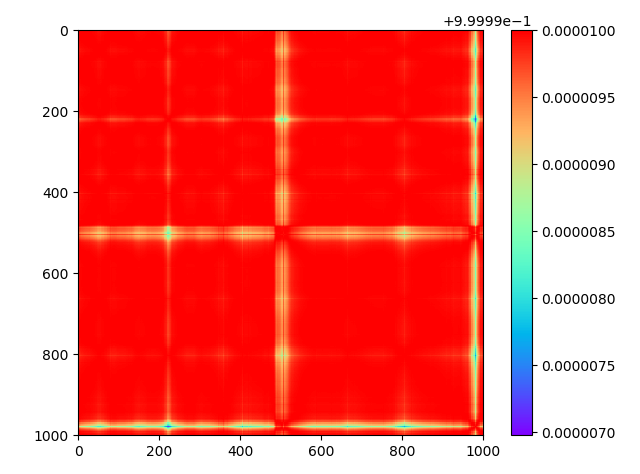}}
\subfigure[\label{figure label}]{\includegraphics[width=0.7\columnwidth]{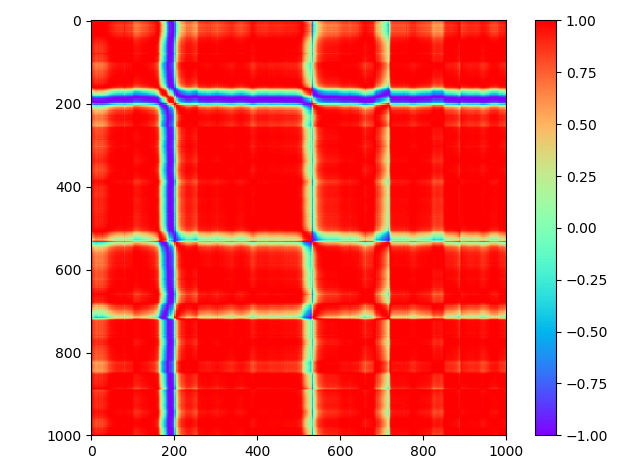}}
\subfigure[\label{figure label}]{\includegraphics[width=0.7\columnwidth]{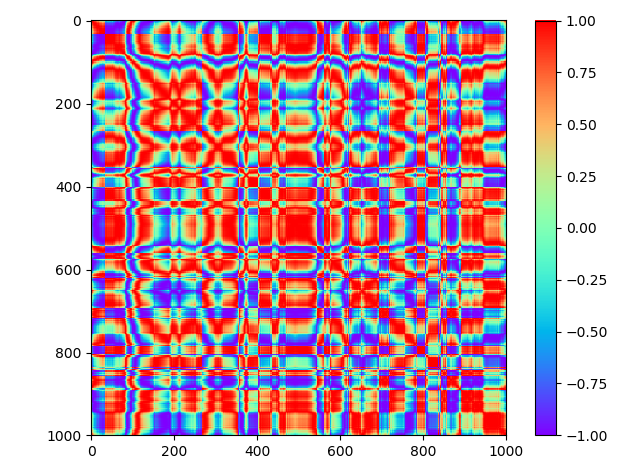}}
\caption{(Color online) Stationary density plot of the correlation matrix elements corresponding to a single realization  of the excitatory-inhibitory model (Eq.\eqref{ei2}), with $Q=1.0$, in  an SW network, for (a) $p=0.03$, (b) $p=0.05$, (c) $p=0.09$ and (d) $p=0.18$.}
\label{fig6:D-ei}
\end{figure*}
%%%%%%%%%%%%%%%%%%%%%%%%%%%%%%%%%%%%%%%%%%%%%%%%%%%%%%%%%%

\subsection{Small-World network}

The time to reach a  stationary state in an SW network (Watts-Strogatz network with rewiring probability $0.03$) is an order of magnitude larger than those of Random networks for both models and could as large as $6\times 10^5$ time steps. 

Figures~\ref{fig3:r-cc} and \ref{fig4:r-ei} illustrate The variations of the total and partial order parameters in terms of the fraction of contrarian and inhibitors, for the conformist-contrarian and excitatory-inhibitory models, respectively. 
For both models, we adapt $Q=0.5, 1.0, 3.0$, and each data point is obtained by averaging over independent random initial conditions. In contrast to the random network, the averaged order parameters are less than unity for $p=0$, which is due to the formation of the defect patterns in small-world networks \cite{esfahani2012noise}. Indeed, for a group of identical phase-oscillators, depending on initial phase distribution, the magnitude of stationary order parameters varies between 0 and 1, then its average is less than 1. The oscillators in the center of defects are in a $\pi$-locked state with the ones in the homogeneous parts, indicating that the Kuramoto dynamics turns some individual oscillators to contrarians even for an identical ensemble of phase oscillators.    

Figure~\ref{fig3:r-cc} shows that the behavior of order parameters versus the fraction of contrarians does not depend on $Q$ and like the random network, the exact overlap of the total order parameter and the relation $r_{\infty}=(1-p) r_{\rm conf}-p r_{\rm cont}$, indicates that the phase difference between conformist and contrarians are always equal to $\pi$; hence the system is in blurred $\pi$-state for all values of $p$ and $Q$. Moreover, Figure~\ref{fig3:r-cc} represents the enhancement of synchronization by increasing the fraction of contrarians to $p\sim 0.04$, at which the synchrony reaches a maximum and then begins to fall by further increasing of $p$. 

For the excitatory-inhibitory model, as can be seen in figure~\ref{fig4:r-ei}, the relation $r_{\infty}=(1-p) r_{\rm excit}+p r_{\rm inhib}$ fits very well with the total order parameter, meaning that the excitatory and inhibitory oscillators, regardless of the value of $p$ and $Q$, are always in-phase. This model also shows a maximum for the total order parameter; however, the optimal value of $p$, which leads to this maximum synchrony, decreases by increasing the strength of inhibitory links to the excitatory ones. 

The non-monotonic behavior of order parameters versus  $p$ can be explained by weakening and reducing the defects as the contrarians or inhibitors are introduced to the dynamics. Figures~\ref{fig5:D-cc} and \ref{fig6:D-ei}, respectively represent the density plots of the elements of the stationary correlation matrix for a single realization of the conformist-contrarian and excitatory-inhibitory models with $Q=1$.  
Figure~\ref{fig5:D-cc} shows the reduction of the isolated defects form 3 to 2 and their weakening in going from $p=0.03$ to $0.05$, for the conformist-contrarian model. Remarkable weakening of defects can also be seen in figure~\ref{fig6:D-ei} for the excitatory-inhibitory model, when the fraction of inhibitors rises from $p=0.03$ to $0.05$.   

To make sure that theses results are not the artifact of using only a fixed network realization, we performed the above simulations on $10$ realizations of SW network with the same number of nodes, degree, and rewiring probability, but with different adjacency matrices and observed similar results in all of them.

As a result, when several contrarians and inhibitors are randomly substituted in the network, some sit in the vicinity of defect locations.  The perturbing effect of these new agents on the oscillators inside the defects, which are in an anti-phase state with other oscillators of their group, give them more freedom to deviate from their previous anti-phase state and so lead to the weakening if the defects. When the fraction of contrarians or inhibitors reaches an optimal value, they give maximal freedom to the defects and lift the synchronization to a maximum. As explained before, the optimal value of $p$ is independent of $ Q $ for the conformist-contrarian model; however, in the excitatory-inhibitory model, increasing the inhibitory links' weight gives rise to a more substantial perturbation, therefore tend to decrease the optimal $p$.

\section{conclusion}

In summary, we numerically investigated the Kuramoto model with two groups of conformist-contrarian and excitatory-inhibitory phase oscillators, with the same intrinsic frequency, in the sparse random and Watts-Strogatz SW networks. 

In random networks, the conformist-contrarian model finds a stationary state with $\pi$-state that crosses over to blurred $\pi$-state by increasing the fraction of contrarians. On the other hand, for the SW network, the system is always in a blurred $\pi$-state where both conformists and contrarians are partially synchronized. Unlike the all-to-all network, the traveling wave state does not appear in both networks; however, by increasing the fraction of contrarians, the synchronization increases and reaches a maximum at an optimal fraction of contrarians. This counterintuitive observation results from conformist defects' attenuation by introducing the contrarians, which elevate society's consensus up to an optimal number of contrarians. 

We found that the excitatory and inhibitory oscillators always are in-phase in both network types for the excitatory-inhibitory model. We also observed the enhancement of synchrony by increasing the fraction of inhibitors in the SW network. However, unlike the conformist-contrarian model, the fraction of inhibitors at which the order parameters reach a peak decreases by increasing inhibitory links' weight. 

We note that enhancement of synchronization by Contrariety and Inhibition in this model could be considered an implication of {\em asymmetry induced symmetry}~\cite{nishikawa2016symmetric,zhang2017asymmetry,zhang2018identical,motter2018antagonistic}, stating that a symmetric state can be induced in an oscillator system by reducing the symmetry of pairwise interactions. Here, we found that the reduction of symmetry in the couplings of a group of identical oscillators in an SW network pushes it to a more symmetric state by elevating the system's level of synchrony.

%However, while the example of asymmetry-induced synchronization studied by Nishikawa {\it et al.}~\cite{nishikawa2016symmetric}  
%requires the pairwise interactions between the oscillators to be directed, in the model discussed in this work the coupling between each pair of oscillators is undirected (but note that asymmetry induced synchronization has been observed in systems with undirected coupling between oscillators~\cite{zhang2018identical}).

This work provides a simple model, which may help us gain more insight into the constructive role of diversity, appearing in the form of contrarians in human societies and inhibitors in neuronal networks known to have small-world connectivity.

\label{conclusion}

\begin{acknowledgements}
The authors gratefully acknowledge the Sheikh Bahaei National High-Performance Computing Center (SBNHPCC) for providing computing facilities and time. SBNHPCC is supported by the scientific and technological department of presidential office and Isfahan University of Technology (IUT).
\end{acknowledgements}

\bibliographystyle{apsrev4-1}

\bibliography{bibliography-2}

\end{document}